\RequirePackage{lineno}
\documentclass[twocolumn,showpacs,aps,prl,amsmath,amssymb]{revtex4-1}
\usepackage[dvips]{graphicx}
\usepackage{subfigure}

\usepackage{amsmath}
\usepackage{color}
\usepackage{epsfig}
\usepackage{indentfirst}
\usepackage{multirow}
\usepackage{amsmath}
\usepackage{amssymb}
\usepackage{bm}
\usepackage{float}
\usepackage{fancyhdr}
\usepackage{enumerate}

\usepackage[perpage,symbol]{footmisc}
\usepackage[colorlinks,
            linkcolor=blue,
            citecolor=blue,
            urlcolor=blue,
            ]{hyperref}

\usepackage{xspace}
\definecolor{orange}{RGB}{255,165,0}
\definecolor{chocolate}{RGB}{210,105,30}
\definecolor{purple}{RGB}{255,0,255}
\usepackage{epstopdf}
\usepackage{array}
\usepackage{lineno}

\newcommand{\BR}{{\mathcal B}}

\newcommand{\jpsi}{J/\psi}
\newcommand{\pip}{\pi^+}
\newcommand{\pim}{\pi^-}

\newcommand{\ar}{\rightarrow}
\newcommand{\epem}{e^+e^-}

\newcommand{\lclc}{\Lambda_{c}^{+}\overline{\Lambda}_{c}^{-}}
\newcommand{\lc}{\Lambda_{c}^{+}}
\newcommand{\lk}{\Lambda K^{+}}
\newcommand{\lpi}{\Lambda \pi^{+}}

\newcommand{\lctolk}{\Lambda_{c}^{+}\rightarrow\Lambda K^{+}}
\newcommand{\lctolpi}{\Lambda_{c}^{+}\rightarrow\Lambda \pi^{+}}

\begin{document}

\title{\boldmath Measurement of the Branching Fraction of the Singly Cabibbo-Suppressed Decay $\Lambda_{c}^{+}\to\Lambda K^{+}$}

\author
{
\begin{small}
\begin{center}
M.~Ablikim$^{1}$, M.~N.~Achasov$^{11,b}$, P.~Adlarson$^{70}$, M.~Albrecht$^{4}$, R.~Aliberti$^{31}$, A.~Amoroso$^{69A,69C}$, M.~R.~An$^{35}$, Q.~An$^{66,53}$, X.~H.~Bai$^{61}$, Y.~Bai$^{52}$
% ...
\\
\vspace{0.2cm}
(BESIII Collaboration)\\
\vspace{0.2cm} {\it
$^{1}$ Institute of High Energy Physics, Beijing 100049, People's Republic of China\\
% ...
$^{4}$ Bochum Ruhr-University, D-44780 Bochum, Germany\\
% ...
$^{11}$ G.I. Budker Institute of Nuclear Physics SB RAS (BINP), Novosibirsk 630090, Russia\\
% ...
$^{31}$ Johannes Gutenberg University of Mainz, Johann-Joachim-Becher-Weg 45, D-55099 Mainz, Germany\\
% ...
$^{35}$ Liaoning Normal University, Dalian 116029, People's Republic of China\\
% ...
$^{52}$ Southeast University, Nanjing 211100, People's Republic of China\\
$^{53}$ State Key Laboratory of Particle Detection and Electronics, Beijing 100049, Hefei 230026, People's Republic of China\\
% ...
$^{61}$ University of Jinan, Jinan 250022, People's Republic of China\\
% ...
$^{66}$ University of Science and Technology of China, Hefei 230026, People's Republic of China\\
% ...
$^{69}$ University of Turin and INFN, (A)University of Turin, I-10125, Turin, Italy; (B)University of Eastern Piedmont, I-15121, Alessandria, Italy; (C)INFN, I-10125, Turin, Italy\\
$^{70}$ Uppsala University, Box 516, SE-75120 Uppsala, Sweden\\
% ...
$^{b}$ Also at the Novosibirsk State University, Novosibirsk, 630090, Russia\\
% ...
}\end{center}
\vspace{0.4cm}
\end{small}
}

\date{\today}


\begin{abstract}
We report a branching fraction measurement of the singly Cabibbo-suppressed decay $\Lambda_{c}^{+}\to\Lambda K^{+}$
using a data sample collected with the BESIII detector at the BEPCII storage ring.
The data span center-of-mass energies from 4.599 to 4.950 GeV and correspond to an integrated luminosity of 6.44 fb$^{-1}$.
The branching fraction of $\Lambda_{c}^{+}\to\Lambda K^{+}$ relative to that of the Cabibbo-favored decay $\Lambda_{c}^{+}\to\Lambda \pi^{+}$ is measured to be
$\mathcal{R}=\frac{\BR(\Lambda_{c}^{+}\to\Lambda K^{+})}{\BR(\Lambda_{c}^{+}\to\Lambda \pi^{+})}=(4.78\pm0.34\pm0.20)\%$. Combining with the world-average value of
$\BR(\Lambda_{c}^{+}\to\Lambda \pi^{+})$, we obtain $\mathcal{B}(\Lambda_c^+\to\Lambda K^+)=(6.21\pm0.44\pm0.26\pm0.34)\times 10^{-4}$.
Here the first uncertainties are statistical, the second systematic, and the third comes from the uncertainty of the $\Lambda_{c}^{+}\to\Lambda \pi^{+}$ branching fraction.
This result, which is more precise than previous measurements, does not agree with theoretical predictions,
and suggests that non-factorizable contributions have been under-estimated in current  models.

\end{abstract}


\maketitle

Since its discovery, there has been continuous interest in understanding the nature of the $\lc$ charmed baryon~\cite{ref3}. Composed of three different quarks, the $\lc$ system is more complicated
than the charmed-meson case and shows a different behavior in both lifetime and decays~\cite{pdg}.
As the lowest-lying charmed baryon state, typical decays of the $\lc$ involve the weak interaction.
Unlike for the case of charmed-meson decays where the factorizable contributions are dominant
due to the large amount of emitted energy~\cite{kamal}, the hadronic weak decays of the $\lc$ are neither color nor helicity
suppressed~\cite{ref0}, and are thus subject to sizeable non-factorizable contributions, such as $W$-exchange diagrams.
This phenomenon is observed in
recent experimental studies of the decays $\lc\ar\Sigma^0\pi^+$,  $\lc\ar\Sigma^+\pi^0$~\cite{ref1} and $\lc\ar\Xi^0 K^+$~\cite{ref2},
which indicate that non-factorizable contributions are important.

To effectively describe the hadronic weak decay of the $\lc$ baryon,
theoretical approaches such as current algebra~\cite{current}, $SU(3)$ flavor symmetry~\cite{CB1,su3} {\it etc.}~are employed to calculate the decay rates.
However, it is challenging to directly evaluate the non-factorizable
decay amplitudes in a model-independent manner, and so the theoretical predictions
rely on phenomenological models.
Experimentally, progress in the investigations of $\lc$ decay
has been relatively slow due to the lack of experimental data in recent decades,
especially for Cabibbo-suppressed decays whose branching fractions are usually smaller than $10^{-3}$. Therefore, further
precise measurements of the branching fractions of $\lc$ hadronic
weak decays are eagerly sought in order to confront theory. Moreover,
experimental measurements can also be taken as input to constrain
these phenomenological models, as they quantify the non-factorizable effects,
and thus will help to improve our understanding of the dynamics of charmed baryons.

\begin{figure}[!htp]
 \begin{center}
     \includegraphics[width=0.23\textwidth]{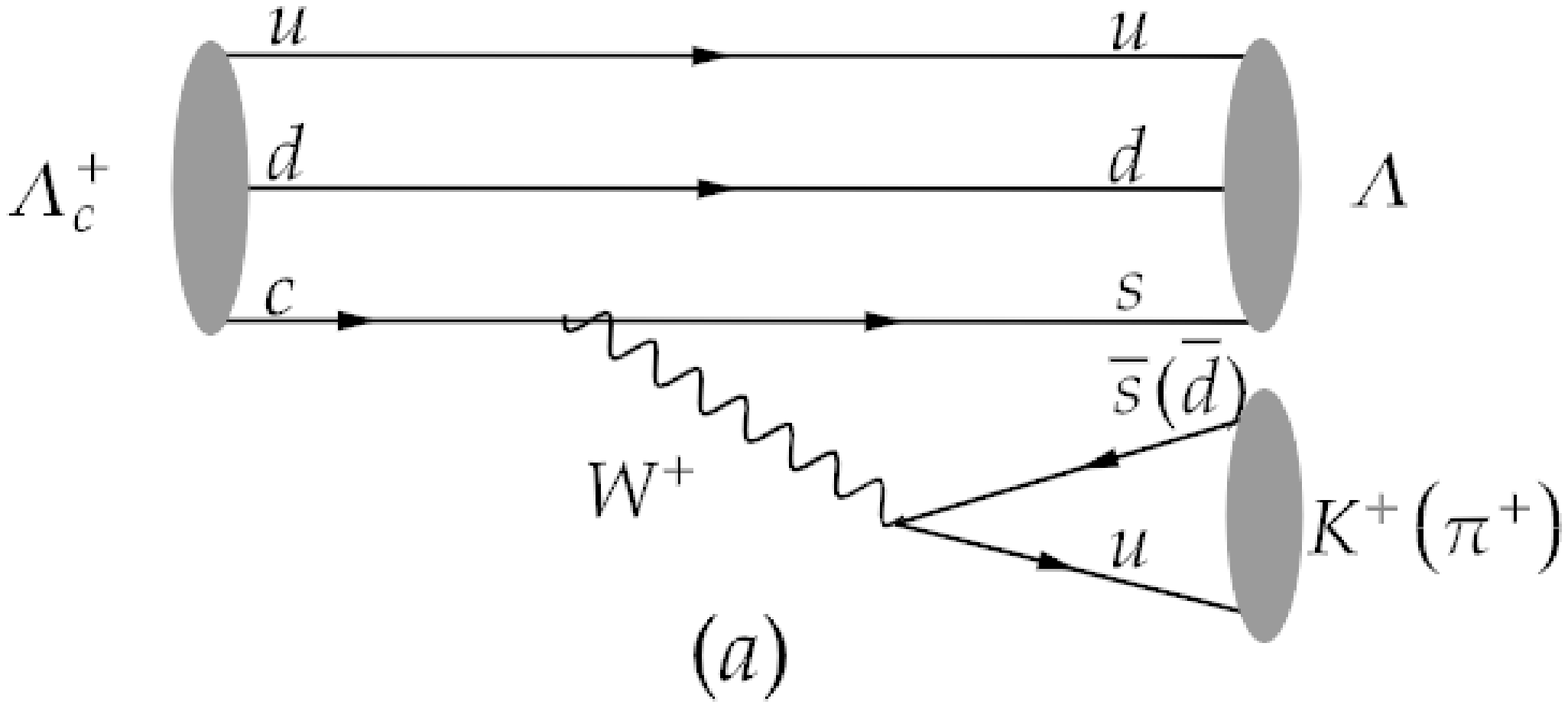}
     \includegraphics[width=0.23\textwidth]{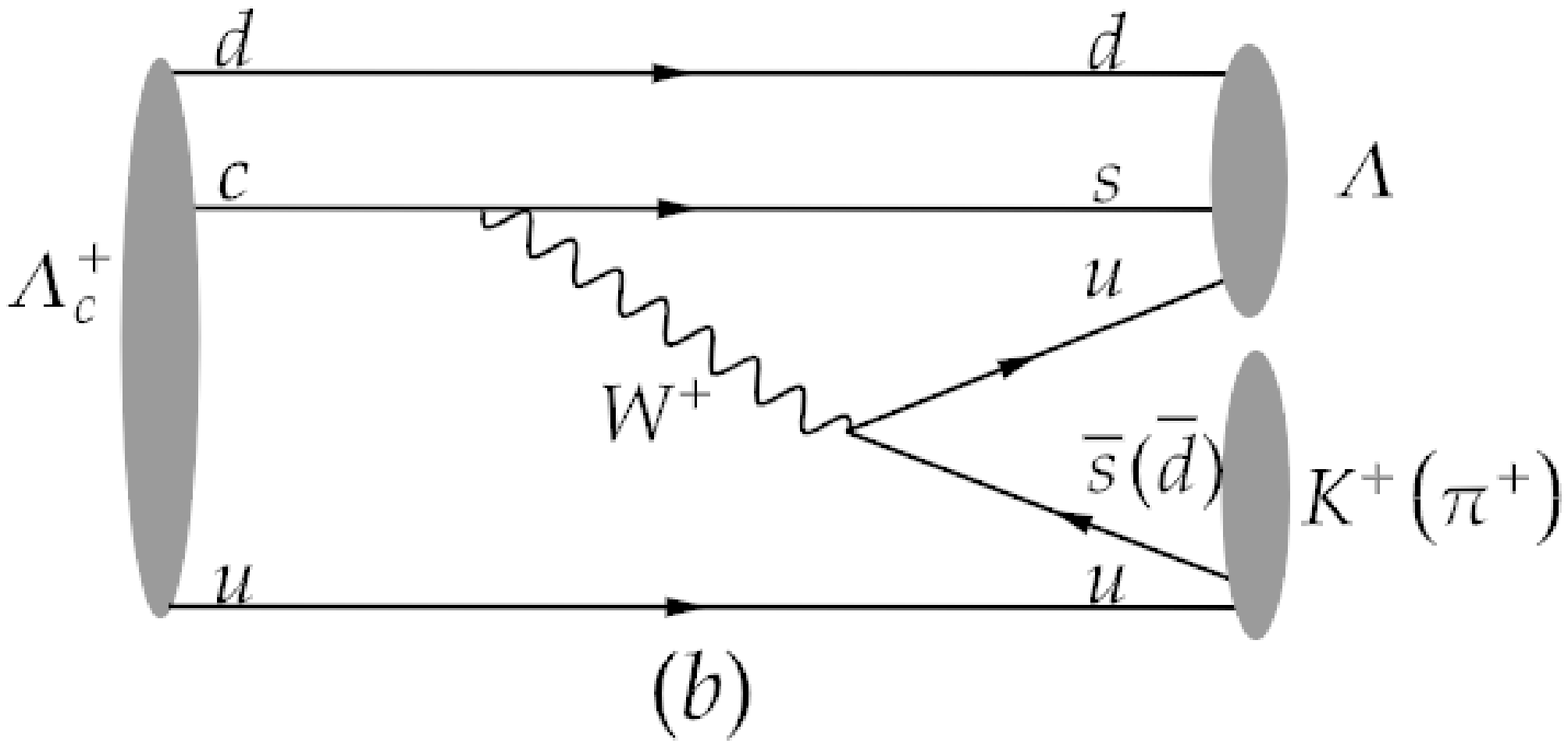}
     \includegraphics[width=0.23\textwidth]{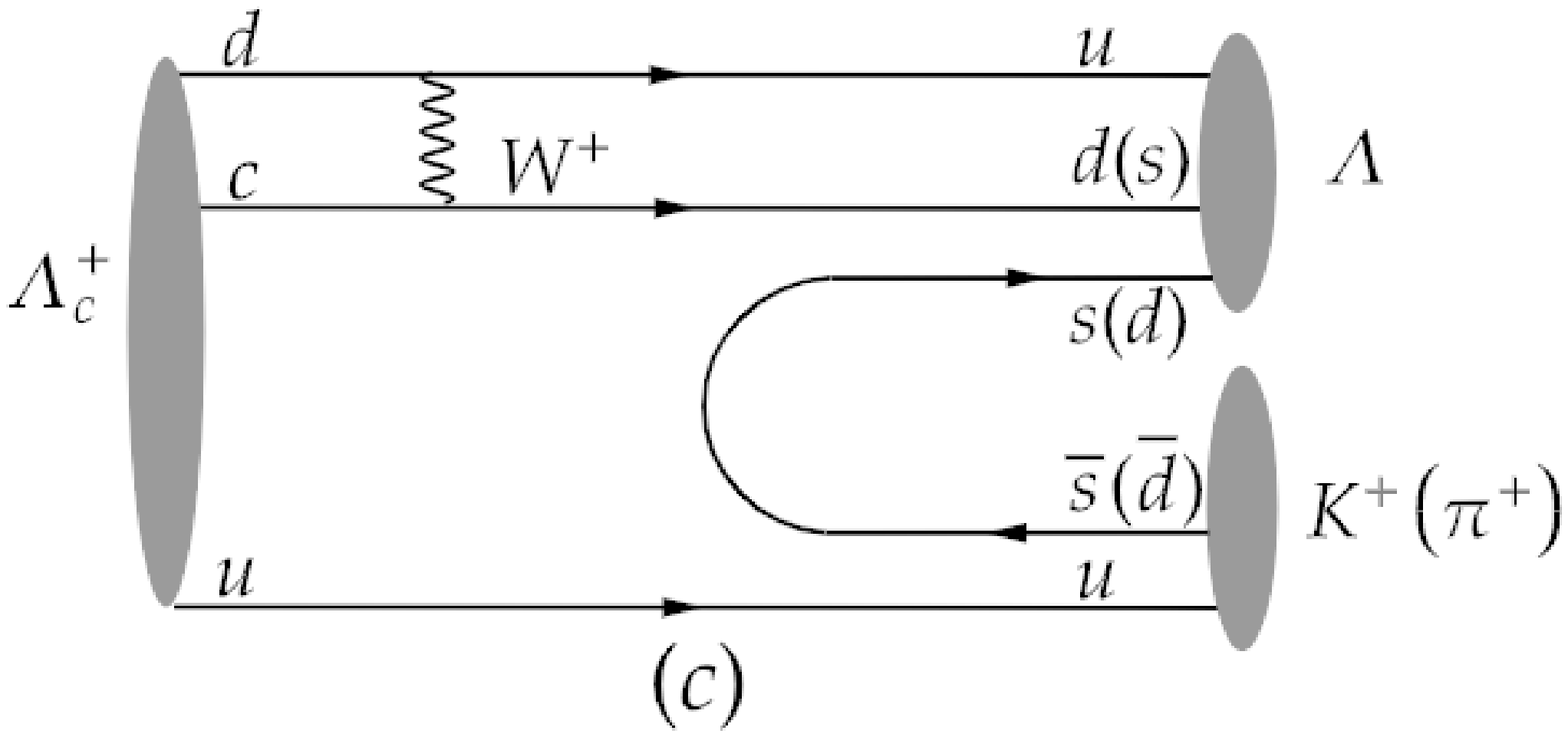}
     \includegraphics[width=0.23\textwidth]{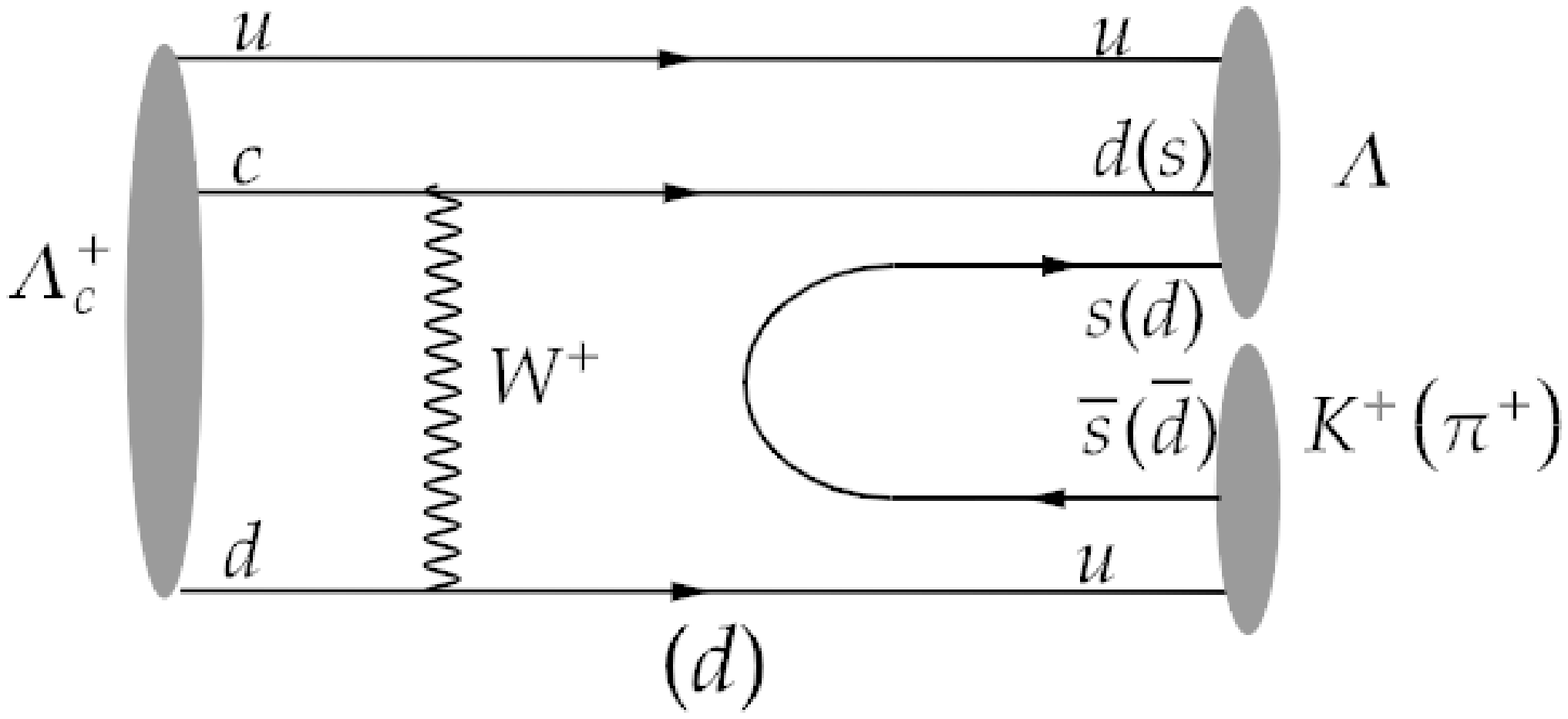}
     \caption{The (a) external emission, (b) internal emission, and (c)(d) $W$-exchange Feynman diagrams for $\Lambda_{c}^{+}\ar\Lambda K^{+}$ and $\Lambda_{c}^{+}\ar\lpi$.}
    \label{feymann}
 \end{center}
 \end{figure}

The singly Cabibbo-suppressed decay $\lc\ar\Lambda K^{+}$ was first studied by the Belle~\cite{ref5}
and BaBar~\cite{ref6} Collaborations more than 15 years ago.
Belle measured the branching fraction of
$\lctolk$ relative to $\lctolpi$ to be $\mathcal{R}=\frac{\BR(\lc\ar\lk)}{\BR (\lc\ar\lpi)}=(7.4\pm1.0\pm1.2)\%$,
while BaBar  reported $\mathcal{R}=(4.4\pm0.4\pm0.3)\%$.
These two results differ from each other by around $2\sigma$.
Figure~\ref{feymann} shows the tree-level Feynman diagrams
for $\Lambda_{c}^{+}\ar\Lambda K^{+}(\pip)$. The contribution from penguin diagrams are 6-orders of magnitude lower
and are thus ignored here~\cite{ref7}.
The external-emission diagram shown in Fig.~\ref{feymann}(a)  is factorizable and
contributes $\sim(\tan\theta_c f_K/f_\pi)^2=7.6\%$ to the relative branching fraction $\frac{\BR(\lc\ar\lk)}{\BR (\lc\ar\lpi)}$
(neglecting the mass difference between pion and kaon),
where $\theta_c$ is the Cabibbo-mixing angle and $f_K(f_\pi)$ is the $K(\pi)$ decay constant.
A more detailed calculation that takes into account the $q^2$-dependent $\Lambda_c-\Lambda$ form factors
and $K(\pi)$ mass difference gives the relative decay branching fraction from this factorizable diagram to
be $\mathcal{R}_{\rm fac}=(7.43\pm0.14)\%$~\cite{ref8},
where the uncertainty comes from knowledge of the form factors.
Refs.~\cite{CB1,CB2,CB3} have calculated the branching fraction
of $\lc\ar\lk$ including the non-factorizable contributions of Figs.~\ref{feymann} (b-d), employing different approaches as summarized in Table~\ref{aaa}
(note that the results from Refs.~\cite{CB4,CB5} are not pure predictions and depend on fits to data).

 \begin{table}[!htp]
\centering
\caption{Theoretical predictions on the branching fraction of $\lc\ar\lk$.}
\label{aaa}
\begin{tabular}{cc}
\hline
\hline
 Theoretical predictions   & $\mathcal{B}(\lc\to\lk)$ ($\times 10^{-3}$) \\
\hline
$SU(3)$ flavor symmetry~\cite{CB1}  &1.4\\
Constituent quark model~\cite{CB2}  &1.2\\
Current algebra~\cite{CB3}  &1.06\\
Diquark picture~\cite{CB4}   &0.18 - 0.39\\
$SU(3)$ flavor symmetry~\cite{CB5}   &0.46$\pm$0.09\\
\hline
\hline
\end{tabular}
\end{table}

In this paper, we report an improved measurement of the branching fraction of the singly Cabibbo-suppressed decay
$\lc\ar\lk$ (referred to as the signal mode) relative to the Cabibbo-favored decay $\lc\ar\lpi$ (referred to as the reference mode)
using a single tag (ST) reconstruction method in $e^+e^-\to\lc\overline{\Lambda}_c^-$ production. The $\Lambda$ is reconstructed through the $p\pi^-$ decay.
Throughout this paper, charge conjugation is always implied unless stated explicitly.
The analysis is based on $(6.44\pm0.04)$ fb$^{-1}$ of $\epem$ annihilation data~\cite{lum,lum2} collected at center-of-mass (c.m.) energies
from 4.599 to 4.950 GeV~\cite{ecm,lum2} with the BESIII detector at the BEPCII storage ring.

The BESIII detector~\cite{Ablikim:2009aa} records symmetric $e^+e^-$ collision events
provided by the BEPCII storage ring~\cite{Yu:IPAC2016-TUYA01}, which operates
in the c.m. energy range from 2.0 to 4.95~GeV.
BESIII has collected large data samples in this energy region~\cite{Ablikim:2019hff}.
The cylindrical core of the BESIII detector covers 93\% of the full solid angle and consists of a helium-based
multilayer drift chamber~(MDC), a plastic scintillator time-of-flight
system~(TOF), and a CsI(Tl) electromagnetic calorimeter~(EMC),
which are all enclosed in a superconducting solenoidal magnet
providing a 1.0~T magnetic field. The solenoid is supported by an
octagonal flux-return yoke with resistive plate counter muon
identification modules interleaved with steel.
The charged-particle momentum resolution at $1~{\rm GeV}/c$ is
$0.5\%$, and the d$E$/d$x$ resolution is $6\%$ for electrons
from Bhabha scattering. The EMC measures photon energies with a
resolution of $2.5\%$ ($5\%$) at $1$~GeV in the barrel (end-cap)
region. The time resolution in the TOF barrel region is 68~ps, while
that in the end-cap region is 110~ps. The end-cap TOF
system was upgraded in 2015 using multi-gap resistive plate chamber
technology, providing a time resolution of
60~ps~\cite{etof}.

A {\sc geant4}~\cite{ref9} based Monte Carlo (MC) simulation package, which includes the geometric
description of the BESIII detector and its response, is used to determine the detection efficiency of signal events,
optimize event-selection criteria, and estimate the backgrounds.
The simulation models the beam-energy spread and initial-state radiation (ISR) in $\epem$ annihilations with the {\sc kkmc} generator~\cite{ref10}.
For ``signal MC" samples, we generate $\epem\to\lclc$ MC events with $\lctolk$ and $\lctolpi$, while the $\overline{\Lambda}_c^-$ baryon decays inclusively. The number of signal MC events which are generated at each c.m.~energy corresponds to that of data. 
For the ISR simulation, the production cross section of $\epem\ar\lclc$ measured by BESIII is incorporated into the {\sc kkmc} program, and the helicity angular distribution $\cos\theta_{\lc}$ in the pair-production process $\epem\ar\lclc$ are also taken into account. For the signal (reference) mode $\lc\ar\lk$ ($\lc\ar\lpi$), the decay angular distributions are described with consideration of the decay asymmetry parameters ($\alpha=-0.84$) of the $\lc$ and $\Lambda$ baryons ($\alpha_{-}=0.732,\alpha_{+}=-0.758$)~\cite{pdg,ref101}. To estimate the proportion of background events, MC samples including the production of $\lclc$ pairs, non-$\lclc$ events, $D\bar{D}$ pairs, ISR production of the $\jpsi$ and $\psi(3686)$ states, and the continuum processes are also generated with {\sc kkmc}~\cite{ref10,ref11}. The known decay modes of charmed hadrons are simulated with {\sc evtgen}~\cite{ref12} with branching fractions taken from the Particle Data Group~\cite{ref13}, and the remaining unknown decay modes are simulated with {\sc lundcharm}~\cite{ref14}.

Charged tracks detected in the MDC are required to be within $|\!\cos\theta|<0.93$,
where $\theta$ is defined with respect to the $z$-axis, which is the symmetry axis of the MDC.
The $\Lambda$ candidate is reconstructed from
a pair of oppositely charged tracks, which are identified as proton and pion, respectively.
Particle identification~(PID)~\cite{PID} for charged tracks combines measurements of the energy loss in the MDC~(d$E$/d$x$)
and the flight time in the TOF to evaluate the likelihoods $\mathcal{L}(h)~(h=p,K,\pi)$ for each hadron $h$ hypothesis.
Tracks are identified as protons when the proton hypothesis satisfies the requirements $\mathcal{L}(p)>\mathcal{L}(\pi)$ and $\mathcal{L}(p)>\mathcal{L}(K)$, while the charged pion is required to satisfy $\mathcal{L}(\pi)>\mathcal{L}(K)$.
Due to the relative long lifetime of $\Lambda$, the proton and pion candidates are further constrained to a common secondary decay vertex.
To effectively separate the secondary vertex from the $\epem$ interaction point (IP),
we require the decay length of the $\Lambda$ to be twice larger than its uncertainty.
The mass window for a $\Lambda$ candidate is defined as $1.111<M(p\pi^-)<1.121$~GeV/$c^2$.

For the signal mode (reference mode), a bachelor kaon (pion) candidate which does not originate
from $\Lambda$ decay is also required. Since the bachelor kaon (pion) track comes directly from
the IP, stricter requirements on the track parameters are applied.
The distance of the closest approach to the IP is required to be within 10~cm along the beam direction,
and 1~cm in the plane perpendicular to the beam direction. 
PID is used to separate the signal mode ($\Lambda K^+$) from the reference mode ($\Lambda \pi^+$), {\it i.e.}
the bachelor kaon (pion) candidate is required to satisfy
$\mathcal{L}(K)>\mathcal{L}(\pi)$ [$\mathcal{L}(\pi)>\mathcal{L}(K)$].

The $\Lambda$ and bachelor kaon (pion) candidates are combined to reconstruct the $\lc$ candidates.
Two kinematic variables, the energy difference $\Delta E = E_{\lc}-E_{\rm beam}$ and
the beam-constrained mass $M_{\rm BC}=\sqrt{E^{2}_{\rm beam}/c^{4}-|\vec{p}_{\lc}|^{2}/c^{2}}$,
are used to identify $\lc$ candidates.
Here $E_{\rm beam}$ is the beam energy and $E_{\lc}$ and $\vec{p}_{\lc}$ are the measured energy
and momentum of the $\lc$ candidate in the $e^+e^-$ c.m. frame. When multiple $\lc$ candidates are found in one event, only the one with the minimum
$|\Delta E|$ is retained for further analysis.
A $\lc$ candidate is finally accepted if $-0.009<\Delta E<0.012$~GeV.

By investigating the MC background events with a generic event-type analysis tool, TopoAna~\cite{topo}, we find that the main background in the $\lctolk$ selection comes from  $\lc\to\Lambda e^+\nu_e$  and  $\lc\ar\Sigma^0\pip$ decays.
The background process $\lc\ar\Lambda e^{+}\nu_{e}$  is rejected by requiring the deposited energy in the EMC divided by the momentum in the MDC ($E/p$) to be less than 0.9 for the kaon candidate. This requirement removes about 80\% of background events, with a signal efficiency loss of about 2.7\%, as indicated by MC simulation.
To avoid losing too much signal efficiency, there is no requirement applied to suppress $\Sigma^0\pip$ contamination.
We find that the $\Lambda_{c}^{+}\ar\Sigma^0\pip$ decay, as well as other irreducible $\lc$ decay backgrounds, contribute a smooth component in the
$M_{BC}$ distribution, which can be well simulated by MC events.

\begin{figure}[!htp]
   \centering
   \includegraphics[width=0.5\textwidth]{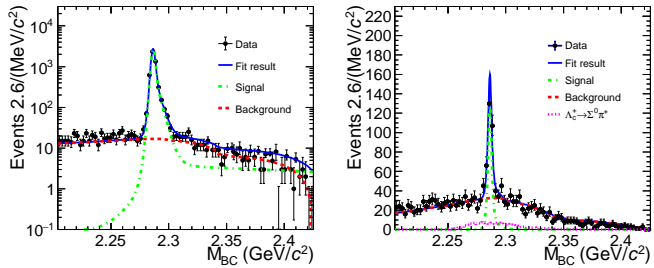}
   \caption{A simultaneous fit to the $M_{\rm BC}$ distributions of the candidates for (left) $\Lambda_c^+\ar\lpi$ and (right) $\Lambda_c^+\ar\lk$. The points with error bars are the full data, the blue solid curves are the sum of fit results at each c.m. energy, the green dot-dashed curves are the signal components, the red dashed curves are the background components, and the magenta dotted curve in the right panel is the normalized $\Sigma^0\pi^+$ background.}
   \label{10}
\end{figure}

Figure~\ref{10} shows the $M_{\rm BC}$ distribution of the accepted candidates for $\lc\ar\lpi$ and $\lc\ar\lk$ from the full data set,
where clear $\lc$ signals can be observed.
To reduce the uncertainty of the $\lctolk$ branching-fraction measurement, we measure the branching fraction of $\lc\ar\lk$ relative to that of $\lc\ar\lpi$. 
A simultaneous fit is performed to the $M_{\rm BC}$ distributions for the data sets at each of the thirteen c.m. energies,
and the signal yield $N_{i}^{\lk}$ for $\lctolk$ events at the $i$-th c.m. energy is further constrained by the relation
$N_{i}^{\lpi} \frac{\epsilon^{\lk}_{i}}{\epsilon^{\lpi}_{i}} \mathcal{R}$,
where $\epsilon^{\lk}_{i}~(\epsilon^{\lpi}_{i})$ is the detection efficiency for the signal (reference) mode,
$N_{i}^{\lpi}$ is the signal yield for the reference mode at the $i$-th c.m. energy,
and $\mathcal{R}=\frac{\BR(\lctolk)}{\BR(\lctolpi)}$ is the relative branching fraction.
The detection efficiencies for the signal and reference modes are estimated by analyzing signal MC events
with the same procedure as for the data analysis, and are listed in Table~\ref{sum}. 
Due to the effects of ISR, the detection efficiencies for  events at $\sqrt{s}>$4.7~GeV
are slightly lower (between 23\% and 29\%),
but the relative efficiency between the signal mode and the reference mode at the
same c.m. energy is quite stable. Thus, these samples, each of which has a  low signal yield, are combined into a merged data set.

%
\begin{table}[H]
\begin{center}
\caption{The integrated luminosity ($\mathcal{L}$) of data set, signal yield ($N$) for $\Lambda\pi^+$ mode from the fit, 
and the detection efficiencies ($\epsilon$ in percentage) for $\lc\ar\lk$ and $\lc\ar\lpi$ modes, respectively at each c.m. energy.
Here the uncertainties are statistical only.}
\label{sum}
\begin{tabular}{c c c  c  c  c }\hline\hline
$\sqrt{s}$~(GeV)& $\mathcal{L}$~(pb$^{-1}$) & $N(\lpi)$   &  $\epsilon(\lpi)$  &	$\epsilon(\lk)$  \\ \hline
 $4.5999$  &$586.9\pm3.9$   &$539.8\pm22.5$                 & $37.9\pm0.2$     &$36.6\pm0.3$   \\
 $4.6118$  &$103.8\pm0.6$   &92.1$\pm$9.4                     & $34.4\pm0.2$     &$33.1\pm0.2$   \\
 $4.6277$  &$521.5\pm2.8$   &502.4$\pm$21.4                & $33.4\pm0.2$     &$32.6\pm0.2$    \\
 $4.6409$  &$552.4\pm3.0$   &$507.8\pm21.6$                & $33.7\pm0.2$     &$31.6\pm0.2$    \\
 $4.6613$  &$529.6\pm2.9$   &$491.4\pm21.1$                & $32.5\pm0.2$     &$30.9\pm0.2$    \\
 $4.6812$  &$1669.3\pm9.0$  &$1470.8\pm36.3$             & $31.4\pm0.2$     &$29.3\pm0.2$    \\
 $4.6984$  &$536.5\pm2.9$   &$374.7\pm18.1$                & $30.2\pm0.2$     &$28.6\pm0.2$    \\
 $> 4.700$ &$1940.1\pm11.6$ &$896.3\pm27.8$              & $24.9$~-~$29.5$     &$23.6$~-~$28.5$    \\
\hline \hline
\end{tabular}
\end{center}
\end{table}

The probability-density functions are constructed with the sum of signal and background components at each c.m. energy. The signal components are modeled with the corresponding MC simulated shapes convolved with Gaussian functions, which account for the resolution difference between data and MC simulation. Here, the standard deviations of the smearing Gaussian resolution function for $\Lambda^+_c\ar\Lambda K^+$ events are constrained to the ones obtained from the fits to $\Lambda^+_c\ar\Lambda \pi^+$ events to improve precision, and the mean values are left as free parameters. The background components are described by ARGUS functions~\cite{ref15} with the truncation parameters fixed to $E_{\rm beam}$ at each c.m. energy.
The simultaneous fit gives
\begin{equation}
\mathcal{R}=\frac{\mathcal{B}(\Lambda_c^+\ar\lk)}{\mathcal{B}(\Lambda_c^+\ar\lpi)}=(4.78\pm0.34)\%,
\end{equation}
where the uncertainty is statistical only. The fit results for the sum of all data sets are shown in Fig.~\ref{10},
where the background curve is the sum of a series of ARGUS functions with a floating endpoint
($E_{\rm beam}$) and showing a complicated distribution.
The corresponding results at each individual c.m. energy are listed in the Appendix.

The main sources of systematic uncertainty on the $\mathcal{R}$ measurement are related to tracking,
PID, $E/p$ requirement, signal shape, and background shape. It should be noted that
many systematic sources, such as those associated with the total number of $\lclc$ events,
$\Lambda$ reconstruction, {\it etc.}, are common to the signal and reference modes and thus cancel in the
$\mathcal{R}$ measurement. In the following, we only discuss the uncorrelated
sources for the signal and the reference modes.

The tracking efficiency of the bachelor $K$ and $\pi$ in the signal and reference modes is not exactly the same due to different momentum distributions, and this leads to an uncertainty in the measurement.
A control sample of $\epem\to K^+K^-\pi^+\pi^-$ is used to study the tracking efficiency of both kaons and pions~\cite{ref16},
and the tracking uncertainties $\delta(p_{T})$ for various transverse momentum intervals are
obtained by comparing the efficiency difference between data and MC simulation.
By assigning each event from the signal MC samples with a corresponding weight [1+$\delta(p_{T})$]
 for both  the signal and reference modes, we re-evaluate the detection efficiencies and find the relative
branching-fraction measurement changes by 1.1\%, which is the systematic uncertainty due to
tracking efficiency.

The PID efficiencies for charged kaons and pions are studied with control samples of $D_s^+\to K^+K^-\pip$,
$D^0\to K^-\pip$, $D^0\to K^-\pip\pip\pim$ decays~\cite{ref16}, and the efficiency
difference $\delta(p)$ between data and MC simulation for different kaon and pion momentum intervals is obtained. A method similar to the one adopted for tracking is applied to assign a systematic uncertainty  of 1.2\% associated with the  PID efficiencies for both the signal and reference modes.

The $E/p$ requirement  introduces a minor efficiency loss for kaons.
The associated systematic uncertainty is assigned to be $0.4\%$ from measuring the efficiency difference between data and MC simulation in  a control sample of  $\Lambda_{c}^{+}\ar p K^{-}\pi^{+}$ decays.

The systematic uncertainty due to the choice of signal shape is studied by fitting the data with an alternative shape,
with free parameters for the smearing Gaussian function of $\Lambda^+_c\ar\Lambda K^+$ mode in the fit.
The change in the signal yield, 0.9\%, is taken as the systematic uncertainty.

To estimate the systematic uncertainty associated with the background shape, we parameterize the background component with an ARGUS
function plus the $\lclc$ inclusive MC shape (accounting for possible unknown $\lc$ decays), or a shape derived from wrong-sign data events.
The largest deviation with respect to the nominal fit result, 2.4\%, is taken as the systematic uncertainty from this source.

The possible systematic bias due to the value of the $\lc\ar\lk$  decay-asymmetry parameters  is studied by considering a range of theoretical predictions for these parameters~\cite{CB1,CB2} as well as a result from the Belle Collaboration~\cite{asymmetry}. The $\lc\ar\lk$ MC samples are re-simulated based on these different 
values and the detection efficiencies are recalculated. The largest deviation with respect to the baseline fit result, 3\%, is assigned as the systematic uncertainty.

Assuming all these sources are independent, the total systematic uncertainty
is calculated to be 4.3\% by adding each contribution in quadrature.

In summary, based on an $\epem$ annihilation data sample of $(6.44\pm0.04)$~fb$^{-1}$
collected at c.m. energies from $\sqrt{s}=4.599$ to 4.950~GeV
with the BESIII detector at the BEPCII storage ring, a study of the singly Cabibbo-suppressed decay
$\Lambda_{c}^{+}\ar\lk$ and the Cabibbo-favored decay $\Lambda_{c}^{+}\ar\lpi$
is performed by using a ST method.
The relative decay branching fraction is measured to be
$\mathcal{R}=(4.78\pm0.34\pm0.20)\%$,
where the first uncertainty is statistical and the second systematic.
Our result is consistent with the measurements performed by the Belle~\cite{ref5} and BaBar~\cite{ref6} Collaborations within uncertainties,
but closer to that of BaBar.
It improves the precision of the PDG average value ($0.047\pm0.009$)~\cite{pdg} 
by a factor of more than two and disfavors theoretical predictions~\cite{CB1,CB2,CB3}. 
By taking the branching fraction of $\mathcal{B}(\Lambda_{c}^{+}\to\Lambda\pi^{+})=(1.30\pm0.07)\%$
as input~\cite{pdg}, we determine $\mathcal{B}(\Lambda_{c}^{+}\to\Lambda K^+)=(6.21\pm0.44\pm0.26\pm0.34)\times 10^{-4}$.

The measured   branching fraction of
$\lc\to\Lambda K^+$ is significantly lower ($\sim40\%$) than the predictions
based on the $SU(3)$ flavor symmetry, constituent quark model, or current algebra~\cite{CB3} listed in Table~\ref{aaa}.
As the  pure factorizable contribution is reliably calculated for the relative branching fraction
($\mathcal{R}_{\rm fac}=(7.43\pm0.14)\%$~\cite{ref8}),we determine the contribution from the non-factorizable effect to be $\mathcal{R}_{\rm non-fac}=\mathcal{R}-\mathcal{R}_{\rm fac}=-(2.65\pm 0.42)\%$, which is negative and has a size comparable to the factorizable contribution.
This indicates that the non-factorizable contributions  in $\lc$ decay
are  important and have been significantly under-estimated in current theoretical models.

It is illustrative to compare our result with analogous ratios measured in different systems.
The ratio of singly Cabibbo-suppressed to Cabibbo-favored decays of $\Lambda_b$ baryons has been measured to be
$\frac{\BR(\Lambda_b\to\lc K^-)}{\BR(\Lambda_b\to\lc\pim)}=(7.31\pm0.16\pm0.16)\%$  by the LHCb Collaboration~\cite{LB-SCS}, which is consistent with the naive expectation $(\tan\theta_c f_K/f_\pi)^2$, and so significantly different for the case with $\Lambda_c$ baryons.
A comparison with 
$\frac{\BR(\lc\to pK^+\pim)}{\BR(\lc\to pK^-\pip)}=(0.82\pm0.12)\tan^4\theta_c$ measured by Belle~\cite{LC-DCS}, shows
that the non-factorizable contribution
in $\lc$ singly Cabibbo-suppressed decay seems to have a more prominent effect.
Compared with $\frac{\BR(D^0\to K^+\pim)}{\BR(D^0\to K^-\pip)}=(1.24\pm0.05)\tan^4\theta_c$ measured by LHCb~\cite{d0} or
$\sqrt{\frac{\BR(D^+\to K^+\pip\pim)}{\BR(D^+\to K^-\pip\pip)}
\frac{\BR(D_s^+\to K^+K^+\pim)}{\BR(D_s^+\to K^+K^-\pip)}}=(1.25\pm0.08)\tan^4\theta_c$ measured by Belle~\cite{d-ds},
our measurement indicates that the $SU(3)$ flavor-symmetry breaking in the charmed
baryon system is more significant than that in the charmed meson case.

The BESIII collaboration thanks the staff of BEPCII and the IHEP computing center for their strong support. This work is supported in part by National Key R\&D Program of China under Contracts Nos. 2020YFA0406400, 2020YFA0406300; National Natural Science Foundation of China (NSFC) under Contracts Nos. 11635010, 11735014, 11835012, 11935015, 11935016, 11935018, 11961141012, 12022510, 12025502, 12035009, 12035013, 12192260, 12192261, 12192262, 12192263, 12192264, 12192265; the Chinese Academy of Sciences (CAS) Large-Scale Scientific Facility Program; Joint Large-Scale Scientific Facility Funds of the NSFC and CAS under Contract No. U1832207; 100 Talents Program of CAS; Project ZR2022JQ02 supported by Shandong Provincial Natural Science Foundation; The Institute of Nuclear and Particle Physics (INPAC) and Shanghai Key Laboratory for Particle Physics and Cosmology; ERC under Contract No. 758462; European Union's Horizon 2020 research and innovation programme under Marie Sklodowska-Curie grant agreement under Contract No. 894790; German Research Foundation DFG under Contracts Nos. 443159800, Collaborative Research Center CRC 1044, GRK 2149; Istituto Nazionale di Fisica Nucleare, Italy; Ministry of Development of Turkey under Contract No. DPT2006K-120470; National Science and Technology fund; National Science Research and Innovation Fund (NSRF) via the Program Management Unit for Human Resources \& Institutional Development, Research and Innovation under Contract No. B16F640076; STFC (United Kingdom); Suranaree University of Technology (SUT), Thailand Science Research and Innovation (TSRI), and National Science Research and Innovation Fund (NSRF) under Contract No. 160355; The Royal Society, UK under Contracts Nos. DH140054, DH160214; The Swedish Research Council; U. S. Department of Energy under Contract No. DE-FG02-05ER41374.

\section*{appendix: SIMULTANEOUS FIT PLOTS FOR THIRTEEN C.M. ENERGY}
\begin{figure*}[]
   \centering
   \includegraphics[width=0.48\textwidth]{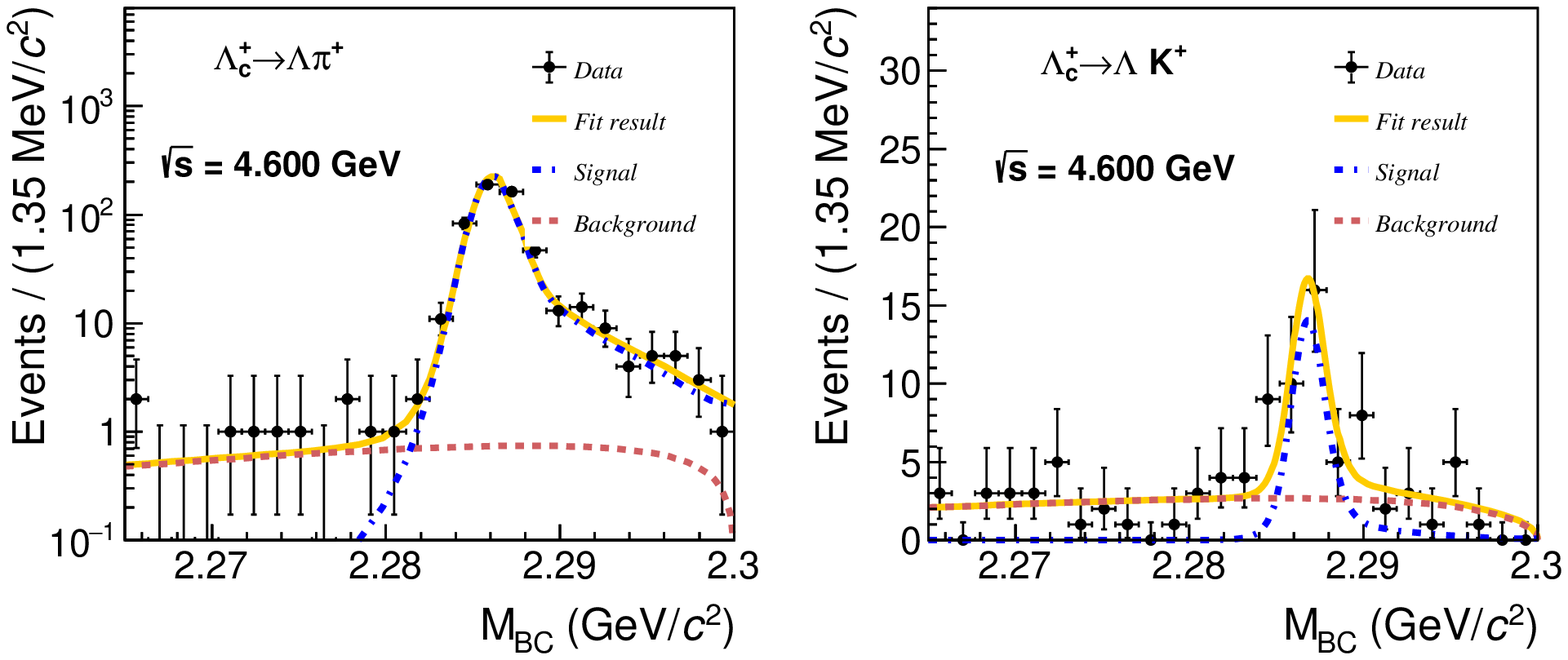}
   \includegraphics[width=0.48\textwidth]{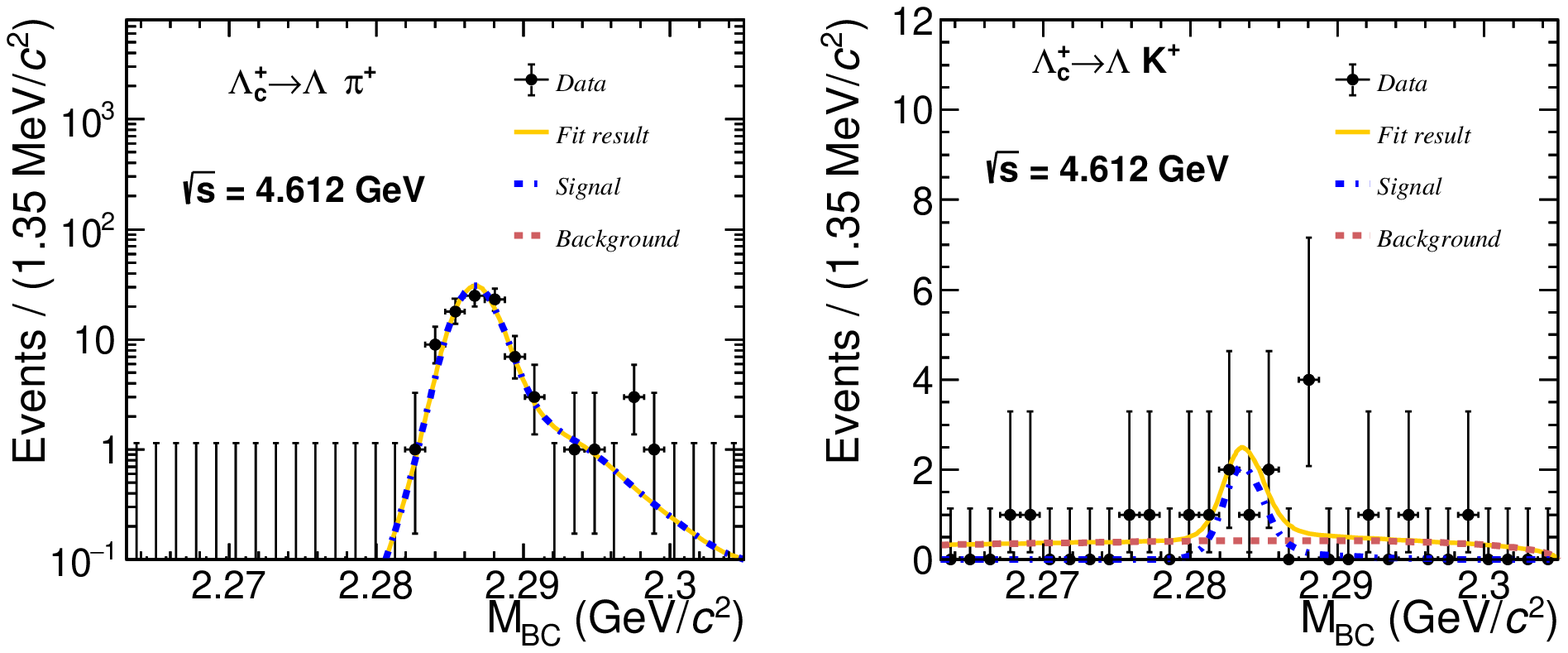}

   \includegraphics[width=0.48\textwidth]{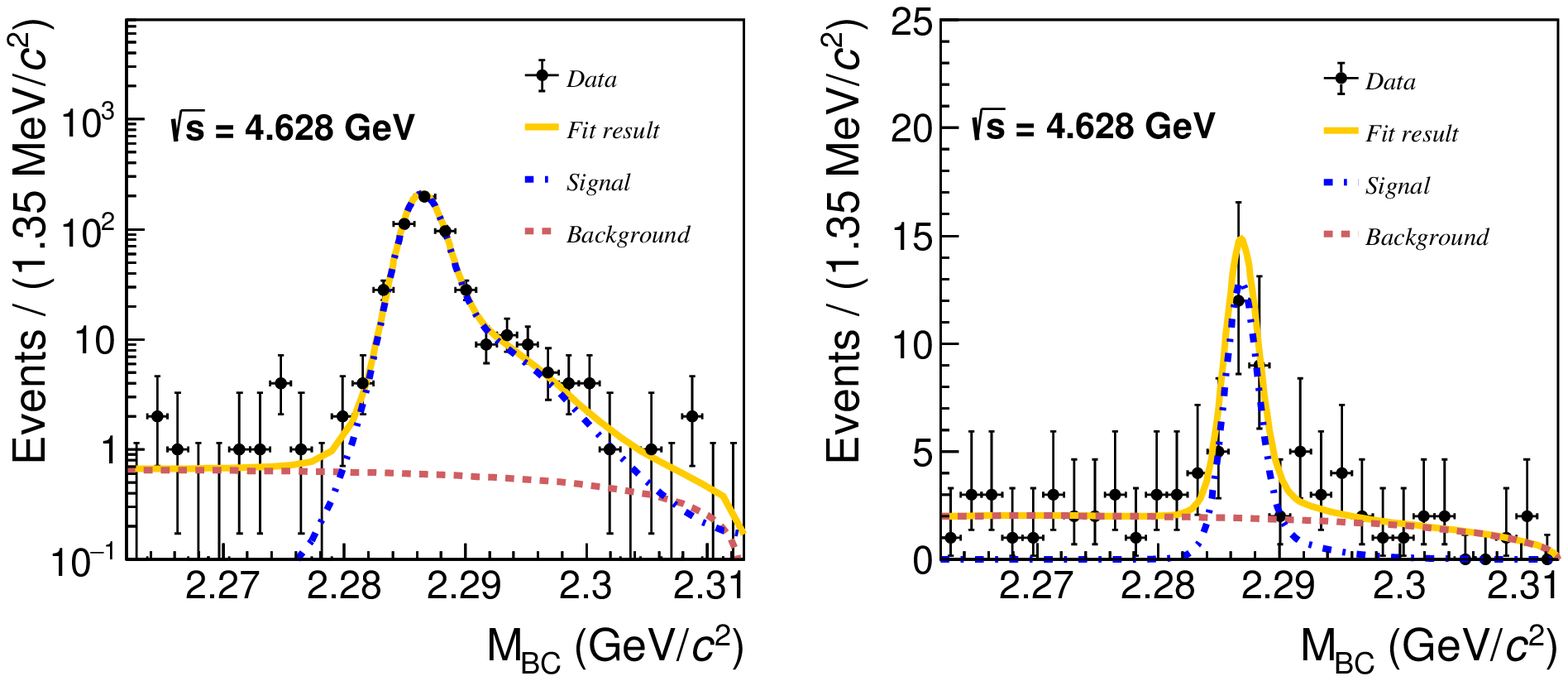}
   \includegraphics[width=0.48\textwidth]{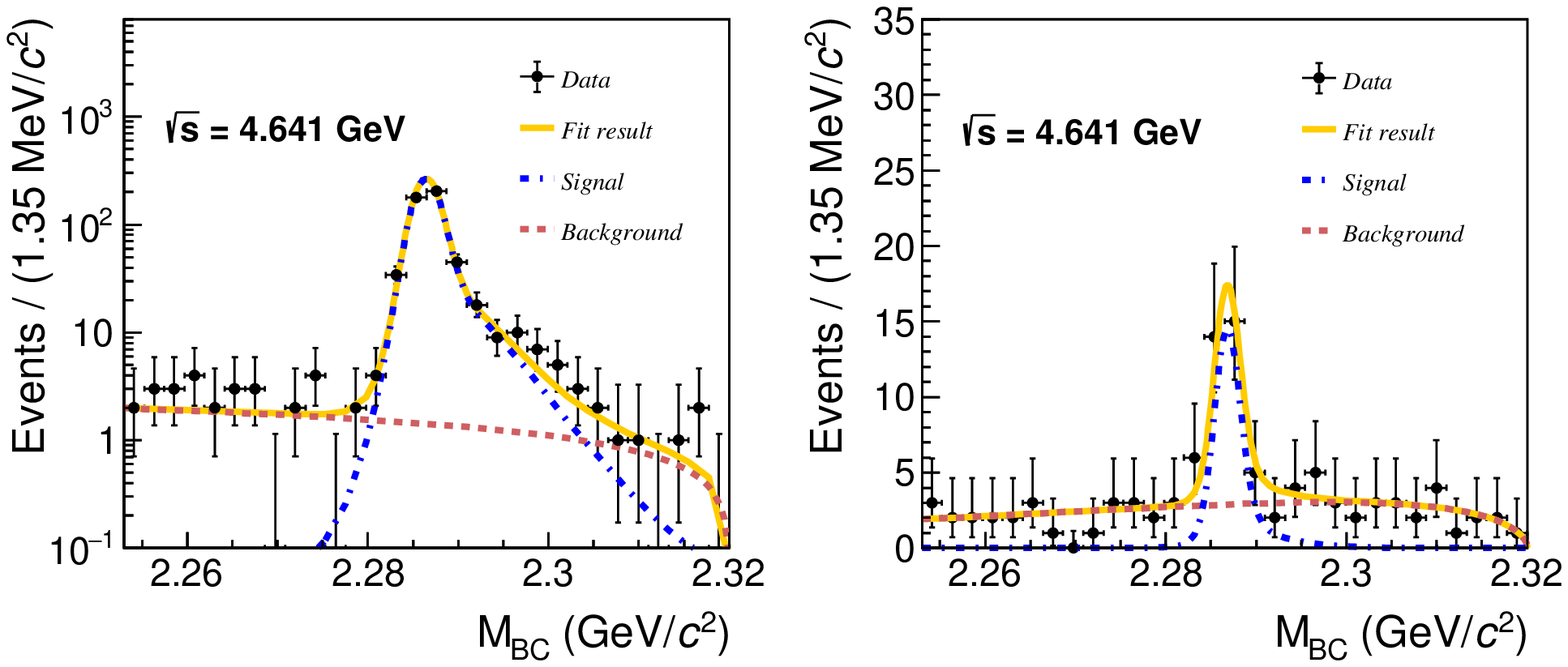}

   \includegraphics[width=0.48\textwidth]{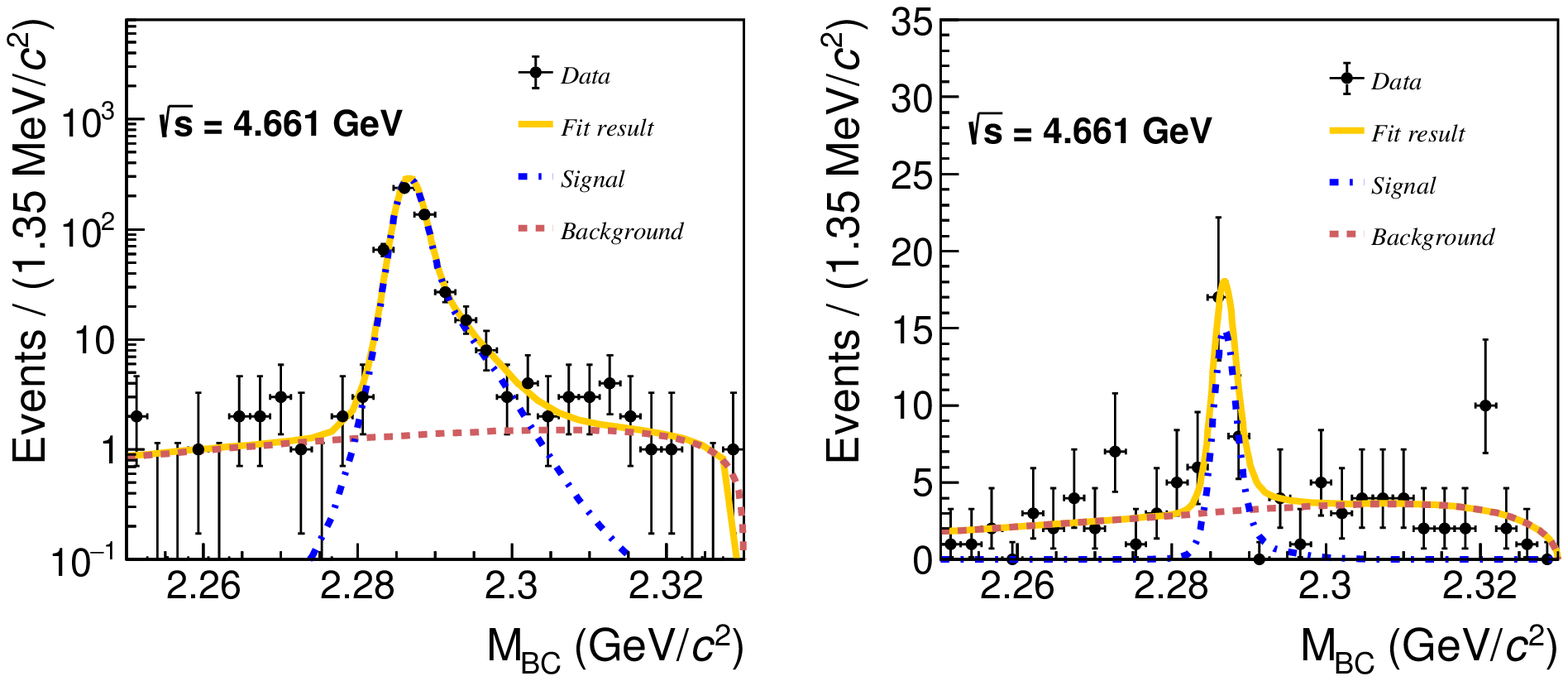}
   \includegraphics[width=0.48\textwidth]{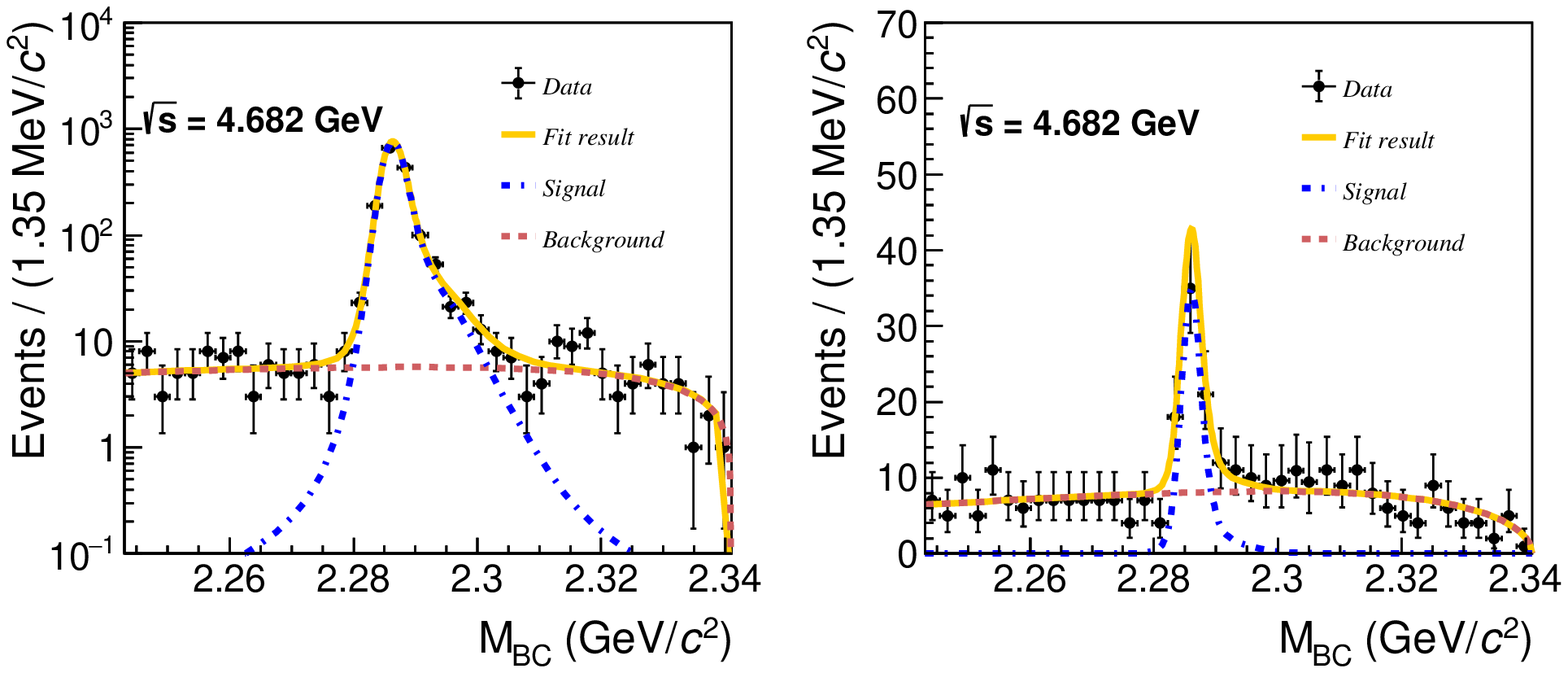}

   \includegraphics[width=0.48\textwidth]{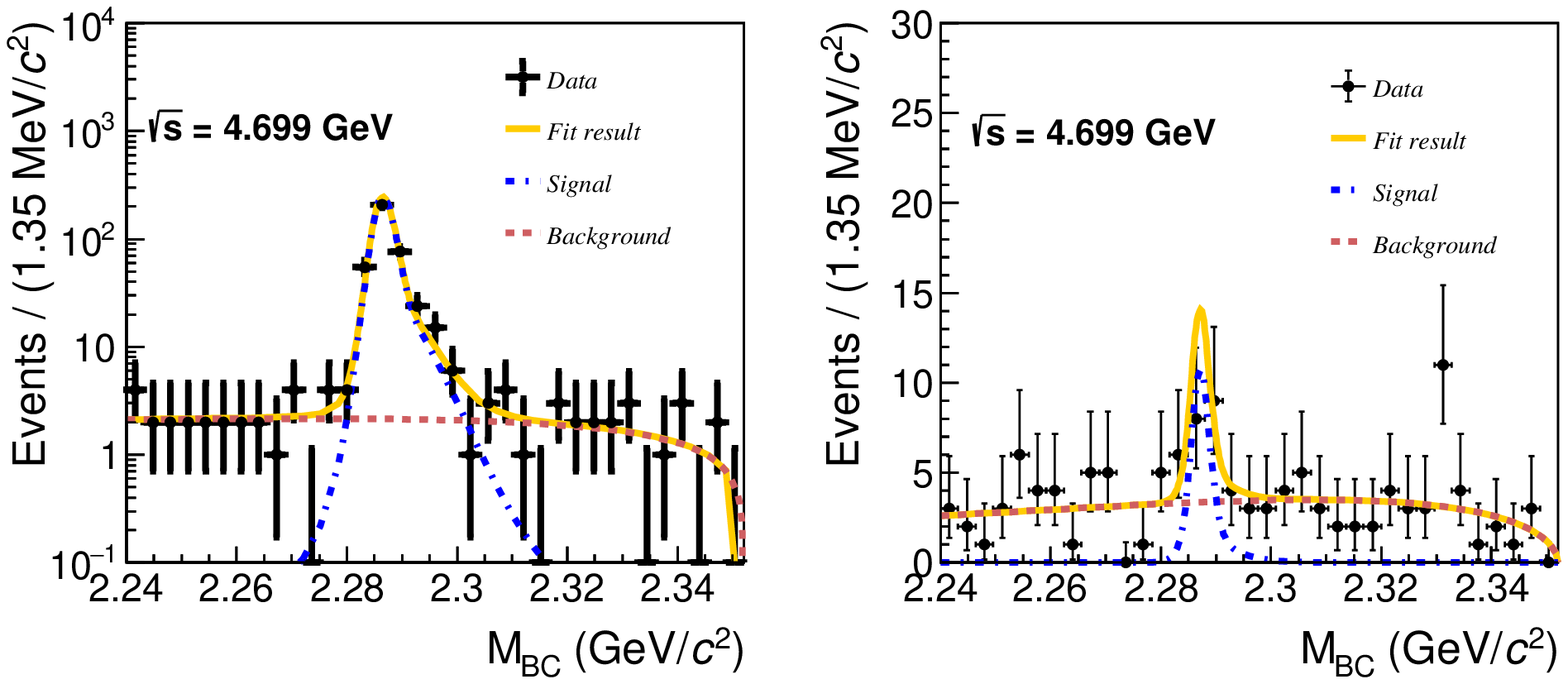}
   \includegraphics[width=0.48\textwidth]{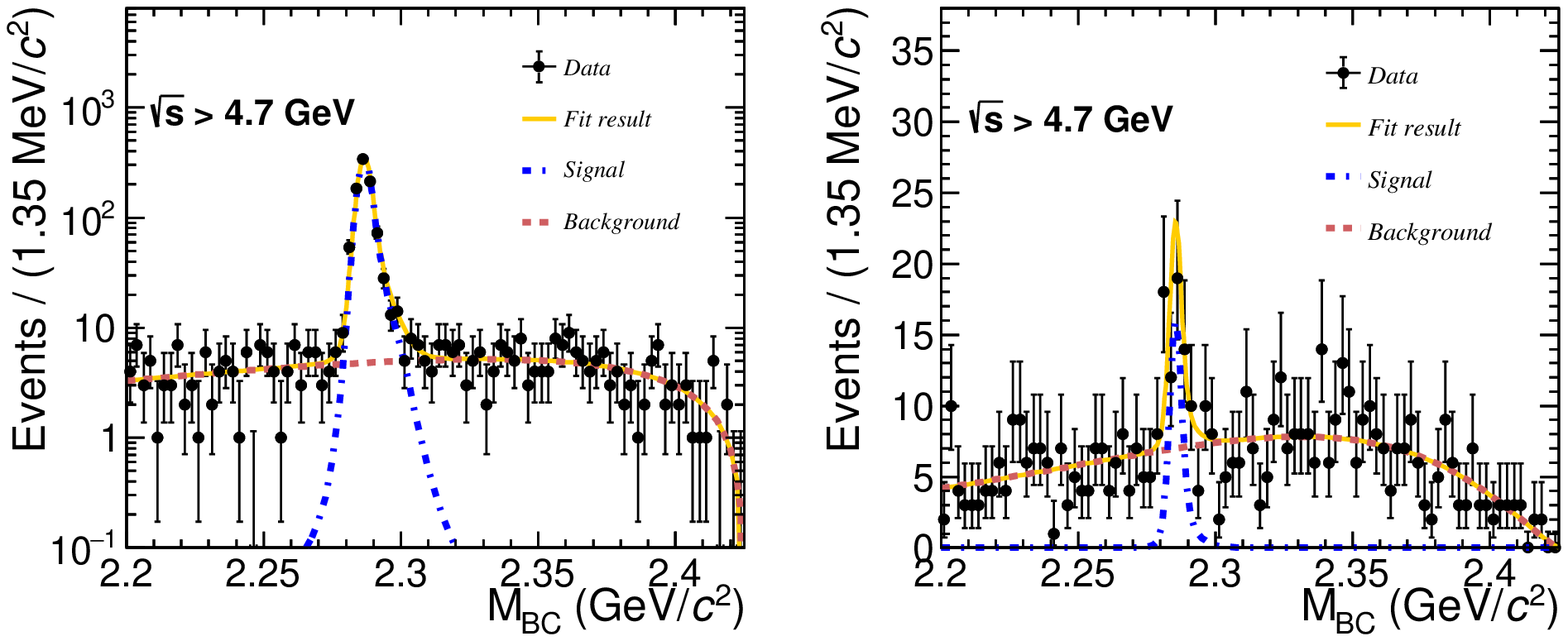}
   \caption{Simultaneous fit result to the $M_{\rm BC}$ distributions of the $\Lambda_c^+\ar\lpi$ (left)
   and $\Lambda_c^+\ar\lk$ (right) candidates at various c.m. energies. The dots with error bars are data, the orange solid curves represent the fit results, the blue dot-dashed curves represent the $\lc$ signal, and the brown dashed curves represent the background components.}
   \label{simfit}
\end{figure*}

Figure~\ref{simfit} shows the fit to the $M_{\rm BC}$ distribution at each c.m. energy.

\end{document}